\documentclass[11p]{article}

\usepackage{amsfonts}
\usepackage{amssymb}
\usepackage{amsmath}
\usepackage{epsfig}

\begin{document}

\title{QUANTUM SHARED BROADCASTING}

\author{V. Giovannetti$^1$ and A. S. Holevo$^2$ \\ \\
{\small $^1$ NEST CNR-INFM \& Scuola Normale Superiore,} \\ {\small Piazza dei Cavalieri
7, I-56126 Pisa, Italy}
 \\ \\
{\small  $^2$ Steklov Mathematical Institute,}
\\ {\small Gubkina 8,
119991 Moscow, Russia}}

\maketitle

\begin{abstract}
A generalization of quantum broadcasting protocol is presented. Here the goal is to copy
an unknown input state into two subsystems which partially overlap.
We show that the possibility of implementing these protocols
strongly depends upon the overlap among the subsystems.
Conditions for approximated shared broadcasting are analyzed.
\end{abstract}

\section{Introduction} \label{sec:intro}
The no-cloning theorem~\cite{NOCLONING} is an important property  of
quantum mechanics which follows from the linear structure of the
theory. In its weaker version the no-cloning theorem formalizes the
physical impossibility of creating a machine that produces exact
copies of an unknown, given quantum state. More generally it
prevents us to construct a machine such that, randomly choosing one
of two non-orthogonal states of a system, it will produce perfect
copies of such state.

Together with entanglement (of which it is a direct consequence)
the no-cloning theorem contributed in
modifying our
approach to information theory and in refining our intuition
about quantum-based information processing~\cite{NIELSEN}
with profound consequences both
in quantum computation and in quantum communication.
In particular it played
an important role in the development of
quantum error correction techniques~\cite{QEC} by preventing one
from having codes that create redundant copies of every
state of a quantum system.

In recent years several generalizations of the no-cloning theorem have
been proposed to include the possibility
of imperfect copies. In particular this yielded
a proof of the impossibility of
cloning with arbitrary high fidelity~\cite{NOCLONING2,CLO1,CLO2,OPTIMALCLO}
an unknown quantum state.
Furthermore, Barnum {\em et al.}~\cite{BCF96}
introduced the idea  of {\em quantum broadcasting} to deal with
mixed input state. Differently from the
original setup~\cite{NOCLONING}, in a broadcasting scenario it is not
required to produce  factorizable copies of the original 
input states. Instead one is allowed to
create a joint (possible entangled) many-body output state composed by
subsystems which locally reproduce the original input state.
In this settings it has been shown that quantum broadcasting is still prohibited
in the case of single input copy and two output copies~\cite{BCF96} but can be done
when starting from a sufficient number of copies~\cite{DMP05}.

No-cloning and quantum broadcasting proved to be an important
investigation tool for characterizing the quantum capacity~\cite{QQ}
of quantum communication channels~\cite{BS}. Using these results in
conjunction with the {\em degradability} properties~\cite{DEVSHOR}
of a quantum communication line one can show that channels which are
{\em anti-degradable}~\cite{CV} must have zero quantum capacity
(see for instance Ref.~\cite{EXAMPLES}).

\begin{figure}[t!]
\begin{center}
\epsfxsize=.9\hsize\leavevmode\epsffile{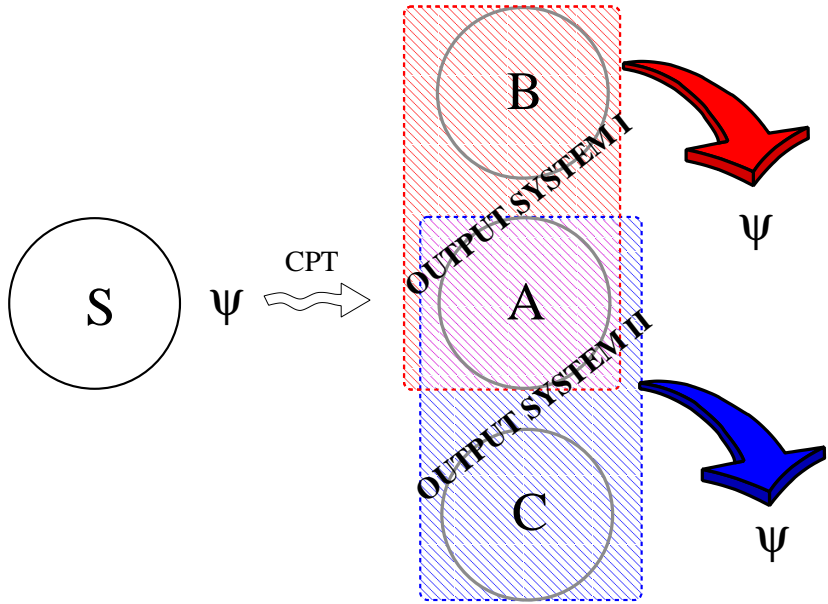}
\end{center}
\caption{Shared broadcasting scheme: states of the source system
$S$ are simultaneously
copied into the output systems I and  II which share a common subsystem $A$. In this
setting  $B$ and $C$  represent the parts of I and II which do not overlap (in other
words I is the joint system $AB$ while II is $AC$).
Standard broadcasting~\cite{BCF96}
is a particular instance of shared broadcasting in
which $A$ is kept into a fix reference state. Perfect shared broadcasting
is possible if and only if $A$ is bigger or equal to the
source space $S$.} \label{fig1}
\end{figure}

In the present paper we would like to discuss a weaker version of
broadcasting that we name {\em Quantum Shared Broadcasting} (QSB).
In its simplest form  the scheme is described in Fig.~\ref{fig1}. As
in the standard broadcasting scenario we have a source system $S$
which provide us with a set of unknown input states $|\psi_S\rangle$
that we would like to duplicate into two output systems I and II. In
the present case however we do not require the output systems to be
independent. Instead we assume a partial {\em overlap} between them.
As shown in the picture, we parameterize such overlap by introducing
a subsystem $A$ which belongs to both the output systems and
representing I and II as the composed systems $AB$ and $AC$
respectively (here $B$ and $C$ are two independent spaces). QSB
succeeds when, for a given unknown input state $|\psi\rangle$ of
$S$, the output density matrix of $ABC$ is such that both its
restrictions
 on I$=AB$ and II$=AC$ contain a copy of $|\psi\rangle$.
As we will see the possibility of realizing such a transformation
strongly depends upon the dimensionality of the shared subsystem
$A$. In particular, perfect shared broadcasting (i.e. a QSB which
produces perfect copies of the states $|\psi\rangle$) cannot be
achieved if $A$ is smaller than the source system $S$, while it is
trivially allowed if $S$ fits in $A$. Similar results apply in the
case of imperfect QSB where one is interested in getting only
approximated copies of the $|\psi\rangle$s: in this case we provide
a threshold which connects the achievable fidelity and the dimension
of $A$.

As in the case of no-cloning, the possibility of  performing
QSB transformations has profound consequences in quantum
communication. In particular this appears to be a fundamental step
for analyzing the communication performances of joint channels.  For
instance QSB can be used to determine whether or not it is
possible to boost the quantum communication efficiency of a given
quantum channel by adding an anti-degradable (zero-quantum capacity)
quantum channel to it. Even though the quantum channel capacity
seems to be a super-additive quantity~\cite{SUPER}, anti-degradable
channels play probably the role of neutral elements in this context.

The paper is organized as follows. We start in
Section~\ref{s:perfect} analyzing the case of perfect broadcasting
showing that this is possible only if the source system $S$ can
fit within the overlap subsystem $A$. In Section~\ref{s:imperfect}
instead we focus on the imperfect broadcasting case providing a
threshold which connects the dimension of $A$ with the fidelity of
the two copies with the input state. The paper then ends with a
perspectives and conclusions section.

%%%%%%%%%%%%%%%%%%%%%%%%%%%%%%%%%%%%%%%%%%%%%%%%%%%
%%%%%%%%%%%%%%%%%%%%%%%%%%%%%%%%%%%%%%%%%%%%%%%%

\section{Perfect Quantum Shared Broadcasting}\label{s:perfect}

Let $S$, $A$, $B$ and $C$ be quantum systems described by the Hilbert spaces ${\cal H}_S$,
${\cal H}_{A}$, ${\cal H}_B$, and ${\cal H}_C$ having dimensions $d_S$, $d_A$, $d_B$ and
$d_C$ respectively. We call $S$ the source system and $A$, $B$, and $C$ the output
systems. Without loss of generality we will assume that $d_S \leqslant
d_A d_B$ and  $d_S \leqslant d_A d_C$. Under these hypothesis  it is always possible to
 find isometries $V_{ABS}$ and $V_{ACS}$ connecting  ${\cal H}_S$ to ${\cal H}_{AB}$ and
${\cal
H}_{AC}$ respectively, that allow us to ``represent''  states of $S$
as states of $AB$ and $AC$.

We are interested in finding a completely positive, trace preserving (CPT) map ${\cal N}$
from $S$ to $A,B,C$ which would allow
us to ``copy'' any vector of ${\cal H}_S$ into $AB$ and $AC$
(see Fig.~\ref{fig1}).
Specifically, given $|{\psi}_S\rangle \in {\cal H}_S$ let us define the density matrix
\begin{eqnarray}
\rho_{ABC} \equiv {\cal N} (|\psi_S\rangle
\langle \psi_S|) \label{xmap}
\end{eqnarray}
with $\rho_{AB} \equiv \mbox{Tr}_C [ \rho_{ABC} ]$ and $\rho_{AC} \equiv
\mbox{Tr}_B [ \rho_{ABC}]$ its reduced density matrices associated with $AB$ and
$AC$.
Our goal is to find ${\cal N}$ and isometries $V_{ABS}:  {\cal H}_S\rightarrow {\cal
H}_{AB}$ and  $V_{ACS}: {\cal H}_S\rightarrow {\cal H}_{AC}$ such that the
fidelities~\cite{FIDELITY}
among the input state $|\psi_S\rangle$
and $\rho_{AB}$, $\rho_{AC}$ are equal to one for all input states $|\psi_S\rangle \in {\cal H}_S$, i.e.
\begin{eqnarray}
\left\{ \begin{array}{l}
F\big( \rho_{AB} ; |\psi_{AB}\rangle \big) \equiv {\langle} \psi_{AB} | \rho_{AB} |
\psi_{AB}\rangle =  1  \\ \\F\big(\rho_{AC} ; |\psi_{AC}\rangle \big) \equiv {\langle} \psi_{AC}
| \rho_{AC}  | \psi_{AC} \rangle =  1   \;,
\end{array}
\right.
\label{afa}
\end{eqnarray}
where
 $| \psi_{AB} \rangle \equiv  V_{ABS} | \psi_S\rangle$ and
$|\psi_{AC}\rangle\equiv V_{ACS} | \psi_S\rangle$ are respectively the ``representations''
of $|\psi_S\rangle$ defined in terms of the isometries   $V_{ABS}$ and $V_{ACS}$. A
channel ${\cal N}$ which satisfies Eq.~(\ref{afa}) for all $|\psi_S\rangle$
of ${\cal H}_S$ is
said to be a perfect Quantum Shared Broadcasting map. Standard broadcasting~\cite{BCF96}
can be obtained from this by
constraining ${\cal N}$ to act trivially on $A$ requiring for instance that
$\rho_{ABC}$ to be of the form $\rho_{BC} \otimes \rho_A$ with $\rho_A$ fixed.

We shall see  that the possibility of achieving shared broadcasting
strongly depends upon the ratio between the dimensions of the source space
$S$ and $A$: if $S$ is small enough to entirely fit inside $A$ then
a perfect QSB map exists, if instead $S$ is bigger than $A$ such a map cannot be
defined.
\newline

{\bf Theorem 1: }
\emph{
Maps ${\cal N}$ which perform perfect QSB
 exist if and only if $A$ is sufficiently big to contain $S$, i.e. if and only if
$d_S \leqslant d_A$. When this happens ${\cal N}$ can
 be chosen to be an isometry.}
\newline

\emph{Proof:}
Showing that ${\cal N}$ exists when $d_S \leqslant d_A$ is trivial.
Indeed when this happens there exists always an isometry $V_{AS}$ which connects
${\cal H}_S$ with ${\cal H}_A$, i.e.
$V_{AS} |\psi_S\rangle=  |\psi_A\rangle$,
for all $|\psi_S\rangle$.
Expand then $V_{AS}$ to construct the isometry
$V_{ABS}$ from ${\cal H}_S$ and ${\cal H}_A \otimes {\cal H}_B$ introduced in
Eq.~(\ref{afa}). This can be done for instance by imposing the conditions
$V_{ABS} |\psi_S\rangle  =   |\psi_A \otimes 0_B\rangle$,
with $|0_B\rangle$
being some reference vector of $B$. Do the same for ${AC}$ by introducing a reference
vector $|0_C\rangle$ on $C$, i.e.
$V_{ACS} |\psi_S\rangle  =   |\psi_A \otimes 0_C\rangle$.
Consider then the transformation which for $|\psi_S\rangle \in {\cal H}_S$ gives
$|\psi_S \rangle \rightarrow |\psi_A \otimes 0_B \otimes 0_C\rangle$.
This is clearly CPT since it is an isometry. Moreover it satisfies the
conditions~(\ref{afa}).
\newline

Let us now consider the case in which $d_S\geqslant d_A + 1$. We
will prove the thesis by contradiction showing that if such ${\cal
N}$ does exist then one can violate the standard no-broadcasting
theorem for vectors belonging to a two-dimensional subspace ${\cal
H}_0$ of ${\cal H}_S$ --- i.e. it would be possible to  construct a
broadcasting machine that creates two perfect copies (one in $B$ and
the other in $C$) of any  vectors of ${\cal H}_0$.

We start observing that the condition (\ref{afa}) implies that the density
matrices $\rho_{AB}$ and $\rho_{AC}$  are pure.
 Specifically, for $X=B,C$  we must have
$\rho_{AX} = |\psi_{AX}\rangle \langle \psi_{AX} |$.
This implies immediately that the global density matrix $\rho_{ABC}$
can be written as
\begin{eqnarray}
\rho_{ABC} = |\psi_{AB}\rangle \langle \psi_{AB} |  \otimes \rho_C
= |\psi_{AC}\rangle \langle \psi_{AC} |  \otimes \rho_B \label{xxx1}\;,
\end{eqnarray}
with $\rho_C$ and $\rho_B$ density matrices which (in principle) may
still depend upon $|\psi_S\rangle$. Take then  the partial trace
with respect to $B$ (or $C$) of both the second and the third term
of Eq.~(\ref{xxx1}). By doing so one arrives to the conclusion
that for all $|\psi_S\rangle$, {\em i)}  $\rho_{ACB}$ must be pure,
 {\em ii)}  $\rho_{ACB}$ must be separable with respect to $A$, $B$ and $C$. That is
\begin{eqnarray}
\rho_{ABC}  = |\phi_A\rangle \langle \phi_A|  \otimes
|\phi_B\rangle \langle \phi_B |  \otimes |\phi_C\rangle \langle
\phi_C |
 \label{xxx2}\;,
\end{eqnarray}
where $|\phi_A \rangle$, $|\phi_B\rangle$ and $|\phi_C\rangle$  are
(not necessarily identical) normalized vectors of ${\cal H}_A$,
${\cal H}_B$ and ${\cal H}_C$ whose dependence upon the input
$|\psi_S\rangle$ will be determined in the following. For $X=B,C$
this gives
\begin{eqnarray}
\rho_{AX}  = |\phi_A\rangle \langle \phi_A |  \otimes
|\phi_X \rangle \langle \phi_X |
 \label{xxx31}\;,
\end{eqnarray}
which, due to Eq.~(\ref{afa}), fixes the action of the isometries
$V_{ABS}$  and $V_{ACS}$ by imposing the conditions
\begin{eqnarray}
V_{AXS} |\psi_S\rangle = |\phi_A \otimes \phi_X \rangle
 \label{xxx3}\;.
\end{eqnarray}
Apply this to an orthonormal basis $\{ |k_S\rangle; k=1,\cdots, d_S\}$ of
${\cal H}_S$, i.e.
\begin{eqnarray}
V_{AXS} | k_S \rangle = |\phi_A^{(k)}  \otimes \phi_X^{(k)} \rangle
 \label{xxx3rrr}\;,
\end{eqnarray}
with $|\phi_A^{(k)}\rangle$ and $|\phi_X^{(k)}\rangle$ defined as in
Eq.~(\ref{xxx3}).
 Since isometries
preserve inner product we must have
\begin{eqnarray}
{\langle} \phi_A^{(k)}   |\phi_A^{(k^\prime)} \rangle  \;
{\langle} \phi_X^{(k)}
|\phi_X^{(k^\prime)}\rangle   = \delta_{kk^\prime}
 \label{xxx5}\;,
\end{eqnarray}
for all $k,k^\prime \in {1,\cdots, d_S}$. Now we remind that the
maximum number of mutually orthonormal vectors in $A$ is $d_A$.
Therefore since $d_S > d_A$, there must exist at least a couple of
$k$, $k^\prime$ (say $k=1$ and $k^\prime =2$) such that ${\langle}
\phi_A^{(1)} |\phi_A^{(2)}\rangle \neq 0$. Consequently
Eq.~(\ref{xxx5}) implies that
\begin{eqnarray}
{\langle} \phi_B^{(1)}|\phi_B^{(2)}\rangle  =  {\langle} \phi_C^{(1)}
|\phi_C^{(2)}\rangle   = 0
 \label{xxx65}\;.
\end{eqnarray}
Consider then the two-dimensional subspace ${\cal H}_0 \subseteq
{\cal H}_S$  generated by $|1_S\rangle$  and $|2_S\rangle$, i.e. the
set of normalized vectors $|\psi_S\rangle = \alpha |1_S\rangle +
\beta |2_S\rangle$. From Eq.~(\ref{xxx3rrr}) and the linearity of
$V_{AXS}$ follows that their ``representations'' on $AX$ are
\begin{eqnarray}
|\psi_{AX}\rangle = V_{AXS} |\psi_S\rangle = \alpha |\phi_A^{(1)}  \otimes
\phi_X^{(1)}\rangle +  \beta |\phi_A^{(2)} \otimes \phi_X^{(2)}\rangle \;.
\nonumber \\
\label{aafa}
\end{eqnarray}
According to Eq.~(\ref{xxx3}) this should be separable with respect to the bipartition
$A-X$. From the orthogonality conditions~(\ref{xxx65}) this is only
possible if
$|\phi_A^{(1)}\rangle  = e^{i \varphi}\;  |\phi_A^{(2)}\rangle$
for some (irrelevant) phase $\varphi$.
Hence for all vectors of ${\cal H}_0$ we can write
\begin{eqnarray}
V_{AXS} |\psi_S\rangle &=&
|\phi_A^{(1)}\rangle  \otimes \Big(\alpha \; |\phi_X^{(1)}\rangle
+   e^{i \varphi} \;\beta \; |\phi_X^{(2)}\rangle   \Big) \nonumber \\
&=&
|\phi_A^{(1)}\rangle \otimes
W_{XS} | \psi_S\rangle  \;,
\label{aafa1}
\end{eqnarray}
with $W_{BS}$  and $W_{CS}$ being isometries  which map ${\cal H}_0$ into
$B$ and $C$ according to the rules
$W_{XS} | 1_S \rangle  =   |\phi_X^{(1)}\rangle$,
$W_{XS} \; | 2_S \rangle  = e^{i \varphi}\;  |\phi_X^{(2)}\rangle$.
Replacing this into Eq.~(\ref{xxx31}) we have finally
\begin{eqnarray}
\rho_{AB}  &=& |\phi_A^{(1)}\rangle \langle \phi_A^{(1)} |  \otimes W_{BS}
|\psi_S\rangle\langle \psi_S | W_{BS}^\dag \nonumber \\ \rho_{AC}  &=&
|\phi_A^{(1)}\rangle \langle \phi_A^{(1)} |  \otimes W_{CS} |\psi_S\rangle\langle \psi_S|
W_{CS}^\dag
 \label{xxx32}\;,
\end{eqnarray}
which shows that the vectors $|\psi_S\rangle \in {\cal H}_0$ have been copied into $B$ and
$C$. This is impossible due to the no-broadcasting theorem. Therefore ${\cal N}$ cannot
exist.~$\blacksquare$

\section{Approximated Quantum Shared Broadcasting} \label{s:imperfect}

A simple generalization of the perfect QSB protocol introduced in the previous section is obtained by
requiring the fidelities among the output copies and the input states
to be higher than a certain fixed threshold:
\newline

{\bf Definition 1:}
{\em  Given ${\varepsilon \in (0,1]}$  we say that $\varepsilon$-QSB from $S$
to $ABC$ is possible
if there exists
a channel ${\cal N}:S\rightarrow ABC$ and
isometries $V_{ABS}:S\rightarrow AB,V_{ACS}:S\rightarrow AC$
such that for an arbitrary input state $|\psi_S\rangle\in{\cal H}_S$  we have}
\begin{eqnarray}
\left\{ \begin{array}{l}
F(\rho_{AB}; |\psi_{AB}\rangle) > 1-\varepsilon \\ \\
F(\rho_{AC}; |\psi_{AC}\rangle) > 1-\varepsilon \;,
\end{array} \right.
\label{afa0000}
\end{eqnarray}
{\em where $\rho_{AB}$, $\rho_{AC}$, $|\psi_{AB}\rangle$, and $|\psi_{AC}\rangle$
are defined as in Eq.~(\ref{afa}).
  We say that Approximated  Quantum Shared Broadcasting (AQSB) from $S$ is possible
if for any $\varepsilon \in (0,1]$ one can find
subsystems $A_\varepsilon \in A$, $B_\varepsilon \in B$, $C_\varepsilon\in C$,
channel ${\cal N}^{(\varepsilon)} : S\rightarrow A_\varepsilon B_\varepsilon C_\varepsilon$ and
isometries $V_{ABS}^{(\varepsilon)}:S\rightarrow A_\varepsilon B_\varepsilon ,
V_{ACS}^{(\varepsilon)}:S\rightarrow A_\varepsilon C_\varepsilon$
which realize an $\varepsilon$-QSB of $S$.
}
\newline

In the following  sections we will give conditions which relate
$\varepsilon$ to $d_A$ and $d_S$ that are necessary  for implementing
$\varepsilon$-QSB protocols and hence AQSB. Before doing so we
notice however that if $d_A<d_S$ \textit{and} the dimensionality of
$BC$ is bounded  then impossibility of AQSB follows from the
impossibility of perfect QSB. This can be shown by contradiction by
noticing that if AQSB would be possible than the sets of channels
${\cal N}^{(\varepsilon)}$ and isometries $V_{ABS}^{(\varepsilon)}$,
$V_{ACS}^{(\varepsilon)}$ would be compact. Therefore letting $\varepsilon
\rightarrow 0$ one will find a limiting channel ${\cal N}^{(0)}$ and
limiting isometries $V_{ABS}^{(0)}$ ,$V_{ACS}^{(0)},$ which fulfill
the perfect QSB impossibility of which was established in
Sec.~\ref{s:perfect}.

\subsection{Notation and preliminary results}\label{s:condtions1}

To deal with the approximations of Eq.~(\ref{afa0000})
we find it useful to review some basic properties of the fidelity
that will be extensively used in the remaining part of the manuscript:

\begin{itemize}
\item[i)] {\em Transitivity:}
Let $\rho$, $\omega$ and $\sigma$ be density matrices, then the
following triangle inequality holds
\begin{eqnarray}
\sqrt{F(\rho;\omega)} \geqslant 1 - \sqrt{ 1- F(\rho; \sigma)} - \sqrt{1-F(\sigma;
\omega)}  \;, \nonumber \\ \label{tri}
\end{eqnarray}
which shows that if $\rho$ and $\omega$ are ``close'' to $\sigma$
then they must be ``close'' to each other too. Equation~(\ref{tri})
can be established by relating $F(\rho; \sigma)$ with the trace
distance $D(\rho;\sigma) \equiv (1/2) \mbox{Tr} |\rho-\sigma|$
through the inequality $1 -\sqrt{F(\rho;\sigma)} \leqslant
D(\rho;\sigma) \leqslant \sqrt{1- F(\rho;\sigma)}$~\cite{NIELSEN}. 
Furthermore Eq.~(\ref{tri}) can be strengthen if at least one
of the density matrices $\rho$ and $\omega$ represents a pure
states. Indeed in this case one gets
\begin{eqnarray}
F(\rho;|\psi\rangle) \geqslant 1 - \sqrt{ 1- F(\rho; \sigma)} - \sqrt{1-F(\sigma;
|\psi\rangle)} \;. \nonumber \\  \label{tripu}
\end{eqnarray}

\item[ii)] {\em Monotonicity under partial trace and purification:} By Bures-Uhlmann theorem~\cite{FIDELITY}
one can easily verify  that the fidelity of the density matrices
 $\rho_{AB}$ and $\sigma_{AB}$ of a joint system $AB$
is always smaller than or equal to the fidelity of the corresponding
reduced density matrices $\rho_A \equiv \mbox{Tr}_B [\rho_{AB}]$ and
$\sigma_A \equiv \mbox{Tr}_B [\sigma_{AB}]$, i.e.
\begin{eqnarray}
F(\rho_{AB}; \sigma_{AB} ) \leqslant F(\rho_A; \sigma_A) \;. \label{fidodido}
\end{eqnarray}
In converse direction, by the same theorem one can also verify that
for any purification $|\varphi_{AB}\rangle$ of  $\rho_A$
there exists
a purification $|\chi_{AB}\rangle$ of $\sigma_A$ such that
\begin{eqnarray}
F(\rho_{A}; \sigma_{A} ) \leqslant F(|\varphi_{AB}\rangle; |\chi_{AB}\rangle)\;. \label{didofido}
\end{eqnarray}

\item[iii)]\emph{Convexity:} Given a density matrix $\rho$ and a vector $|\psi\rangle$,
\begin{eqnarray}
F(\rho; |\psi\rangle) &\leqslant& F(|\phi\rangle;|\psi\rangle) \label{conv1} \;, \\
F(\rho; |\psi\rangle) &\leqslant& \lambda_{\text{max}}\;, \label{conv2}
\end{eqnarray} where $\lambda_{\text{max}}$ is the maximal eigenvalue of $\rho$
and  $|\phi\rangle$ the corresponding eigenvector.
\end{itemize}

Using the above properties it is relatively easy to
generalize the identity~(\ref{xxx2}). Specifically one can show that
if Eq.~(\ref{afa0000}) holds for
all input states $|\psi_S \rangle$, then
there should exist (not necessarily identical)
pure states $|{\phi}_{A}\rangle$, $|\phi_B\rangle$ and $|\phi_C\rangle$ of $A$, $B$, and $C$
such that
the output states $\rho_{ABC}$ of the channel~(\ref{xmap})
are uniformly ``close'' to the tensor product states
$|{\phi}_A \otimes {\phi}_B \otimes
{\phi}_C\rangle$,
while the isometric representations $|\psi_{AB}\rangle$
and $|\psi_{AC}\rangle$ of  $|\psi_S\rangle$ are uniformly ``close'' to
$|{\phi}_{A}\otimes {\phi}_{B} \rangle$
and  $|{\phi}_A \otimes {\phi}_C \rangle$, respectively.
\newline

{\bf Lemma 1:} \emph{If  $\varepsilon$-QSB from $S$ to $ABC$ is possible for some given
value $\varepsilon \in (0,1]$,  then for all input states $|\psi_S\rangle$
one can find $|\phi_A\rangle$,
$|\phi_B\rangle$ and $|\phi_C\rangle$ such that
\begin{eqnarray}
F( \rho_{ABC};
|\phi_A\otimes \phi_B \otimes \phi_C\rangle) &>&  1 - 3 \; \varepsilon^{1/8}\;,
\label{goal1} \\
F( |\psi_{AX}\rangle;|\phi_A\otimes \phi_X\rangle )& > & 1 -  \varepsilon^{\prime}
\label{goal1v} \;,
\end{eqnarray}
where $\rho_{ABC}$ and $|\psi_{AX}\rangle$ are defined as in Eqs.~(\ref{xmap}) and (\ref{afa}),
while   $\varepsilon^{\prime} =   2 \; \varepsilon^{1/2}$ for $X=B$, and
$\varepsilon^{\prime} =  3.4 \; \varepsilon^{1/8}$ for $X=C$.
(the asymmetry among the $B$ and $C$   is a consequence of the relative
freedom one has
in defining the vectors $|\phi_A\rangle$, $|\phi_B\rangle$, and $|\phi_C\rangle$).}
\newline

{\em Proof:}
Purify the output state
$\rho _{ABC}$ to the vector $|\varphi_{ABCE}\rangle$ with $E$ being
an auxiliary system. Since $|\varphi_{ABCE}\rangle$ is purification
of $\rho_{AB}$ and $\rho_{AC}$, then by the condition~(\ref{afa0000})
and by the property ii) of the fidelity
we have
\begin{eqnarray}
F(|\varphi_{ABCE}\rangle; |\psi_{AB}\otimes \psi^{\prime}_{CE}\rangle)  > F(\rho_{AB};
 |\psi_{AB}\rangle)  >
1 -\varepsilon, \nonumber  \\
F(|\varphi_{ABCE}\rangle; |\psi_{AC}\otimes \psi^{\prime}_{BE}\rangle)  > F(\rho_{AC}; |\psi_{AC}\rangle)  >
1 -\varepsilon. \nonumber \\
 \label{AP0}
\end{eqnarray}
for some $|{\psi}_{CE}^{\prime}\rangle$, $|{\psi}_{BE}^{\prime}\rangle$ (this
simply follows from the fact that {\em any} purification of a pure
state is factorizable). By triangle inequality~(\ref{tripu}) and by
monotonicity under partial trace~(\ref{fidodido}) we then obtain
\begin{eqnarray}
&&F(|\psi_{AB}\otimes \psi^{\prime}_{CE}\rangle; |\psi_{AC}\otimes \psi^{\prime}_{BE}\rangle)  >
1 -2 \sqrt{\varepsilon}\label{AP2}\;, \\ \nonumber  \\
&&F(|\psi_{AB} \rangle; \sigma_A\otimes \sigma_B^{\prime})  >  1 -2 \sqrt{\varepsilon} \;,
\label{AP3} \\
&&F(|\psi_{CE}^{\prime} \rangle; \sigma_C \otimes \sigma_E^{\prime})  >  1 -2 \sqrt{\varepsilon} \;,
\label{AP4}
\end{eqnarray}
with $\sigma_A = \mbox{Tr}_C [ |\psi_{AC}\rangle\langle\psi_{AC}|]$,
$\sigma_C  = \mbox{Tr}_A[ |\psi_{AC}\rangle\langle \psi_{AC}|]$,
$\sigma_B^{\prime} = \mbox{Tr}_E [
|\psi_{BE}^{\prime}\rangle\langle\psi_{BE}^{\prime}|]$, and
$\sigma_E^{\prime}  = \mbox{Tr}_B[ |\psi^{\prime}_{BE}\rangle\langle
\psi_{BE}^{\prime}|]$. Let us now define $|\phi_A\rangle$,
$|\phi_B\rangle$, $|\phi_C\rangle$ and $|\phi_E\rangle$  as the
eigenvectors associated with the maximal eigenvalues of the density
matrices
 $\sigma_A$, $\sigma_B$, $\sigma_C^{\prime}$ and $\sigma_E^{\prime}$, respectively.
According to the property~(\ref{conv1})
we then have
\begin{eqnarray}
F(|\psi_{AB} \rangle; |\phi_A\otimes \phi_B\rangle) &>& F(|\psi_{AB} \rangle; \sigma_A\otimes \sigma_B) \nonumber \\
&>& 1 -2 \sqrt{\varepsilon}\;,
\label{AP5} \\
F(|\psi_{CE}^{\prime} \rangle; |\phi_C\otimes \phi_E\rangle)  &>&  F(|\psi_{CE}^{\prime} \rangle; \sigma_C\otimes \sigma_E) \nonumber \\
&>& 1 -2 \sqrt{\varepsilon} \;.
\label{AP55}
\end{eqnarray}
The first one already proves Eq.~(\ref{goal1v}) for $X=B$. To
proceed apply the triangle inequality~(\ref{tripu}) to
Eqs.~(\ref{AP0}) and (\ref{AP5}). This yields
\begin{eqnarray}
F(|\varphi_{ABCE}\rangle; |\phi_A\otimes \phi_B \otimes \psi^{\prime}_{CE}\rangle) &>&
1 -  \Big({\varepsilon}^{1/4} + \sqrt{2}\Big) \varepsilon^{1/4} \nonumber \\
&>& 1 - 2.5 \;
\varepsilon^{1/4} \;.   \label{AP7}
\end{eqnarray}
Again by triangle inequality between~(\ref{AP55}) and (\ref{AP7}) we
get
\begin{eqnarray}
&&F(|\varphi_{ABCE}\rangle; |\phi_A\otimes \phi_B \otimes \phi_{C}\otimes \phi_E\rangle)  \nonumber \\
&&\qquad \qquad \quad \qquad >
1- \Big(\sqrt{2}  \; \varepsilon^{1/8} + \sqrt{ {\varepsilon}^{1/4} + \sqrt{2}  } \Big)
\varepsilon^{1/8} \nonumber \\ &&\qquad\quad\qquad \qquad >  1 - 3.0 \;
\varepsilon^{1/8}\;, \label{AP50}
\end{eqnarray}
which, by partial trace with respect to $E$, gives Eq.~(\ref{goal1}).

To derive the case $X=C$ of Eq.~(\ref{goal1v}) we first apply to
Eqs.~(\ref{AP2}) and (\ref{AP5}) the triangle
inequality~(\ref{tripu}) and then the monotonicity under partial
trace~(\ref{fidodido}), obtaining
\begin{eqnarray}
&&F(|\psi_{AC} \otimes \psi_{BE}^{\prime}\rangle; |\phi_A\otimes\phi_B \otimes \psi_{CE}^{\prime}\rangle )
>   1 -2 \sqrt{2} \; {\varepsilon}^{1/4} \;, \nonumber \\ \label{AP20} \\
&& F(|\psi_{AC}\rangle ; |\phi_A\rangle\langle\phi_A| \otimes \omega_C^{\prime} )
>   1 -2 \sqrt{2} \; {\varepsilon}^{1/4} \;,
\label{AP22}
\end{eqnarray}
with $\omega_C^{\prime} \equiv \mbox{Tr}_E[ |\psi_{CE}^{\prime}\rangle\langle \psi_{CE}^{\prime}|]$.
Taking $|\phi_C^{\prime}\rangle$ the eigenvector of $\omega_C^{\prime}$ associated with its
 maximal
eigenvalue and invoking the property~(\ref{conv1}) we then have,
\begin{eqnarray}
F(|\psi_{AC}\rangle ; |\phi_A\otimes \phi_C^{\prime}\rangle )  &>&
F(|\psi_{AC}\rangle ; |\phi_A\rangle\langle\phi_A| \otimes \omega_C^{\prime} ) \nonumber \\
&>&   1 -2 \sqrt{2} \; {\varepsilon}^{1/4} \;.
\label{AP24}
\end{eqnarray}
By tracing with respect to $A$ we then get
\begin{eqnarray}
F(|\phi_C\rangle; |\phi_C^{\prime}\rangle) >
F(\sigma_C ; | \phi_C^{\prime}\rangle )
>   1 -2 \sqrt{2} \; {\varepsilon}^{1/4} \;,
\label{AP26}
\end{eqnarray}
with $\sigma_C$ and $|\phi_C\rangle$ as in Eq.~(\ref{AP55}). This
together with Eq.~(\ref{AP24}) finally gives
\begin{eqnarray} F(|\psi_{AC}\rangle;
|\phi_A\otimes \phi_C\rangle) &>& 1 - 2 \sqrt{2\sqrt{2}} \; \varepsilon^{1/8}
\nonumber \\
&>& 1 -
3.4 \; \varepsilon^{1/8}\;,
\label{AAA} \end{eqnarray}
which proves Eq.~(\ref{goal1v}) for  $X=C$.
$\blacksquare$

\subsection{Conditions for approximated $\varepsilon$-QSB}\label{s:condtions2}

Starting from the results of the previous section we now
show that for $d_S > d_A$, $\varepsilon$-QSB transformations  cannot be realized
if the parameter $\varepsilon$ is below a certain finite threshold $\varepsilon_0$ that
we have estimated as
\begin{eqnarray}
\varepsilon_0 = \min\big\{ 0.6 \times  10^{-175} ; 2.4 \times 10^{-14} \; d_A^{-8} \big\}
\label{varevare}\;.
\end{eqnarray}
This bound is not optimal and can probably be improved.
However, it  does not
depend upon
the  dimension of the output states $B$ and $C$ and implies that
AQSB transformations are not physical for $d_S > d_A$.
\newline

{\bf Theorem 2: }
\emph{Let  $d_S > d_A$, then it is not possible to have $\varepsilon$-QSB transformations
from $S$ to $ABC$ with $B$ and $C$ generic quantum systems, for
values of  $\varepsilon$ which are smaller than or equal to  $\varepsilon_0$ of
Eq.~(\ref{varevare}).}
\newline

\begin{figure*}
\begin{center}
\epsfxsize=.8\hsize\leavevmode\epsffile{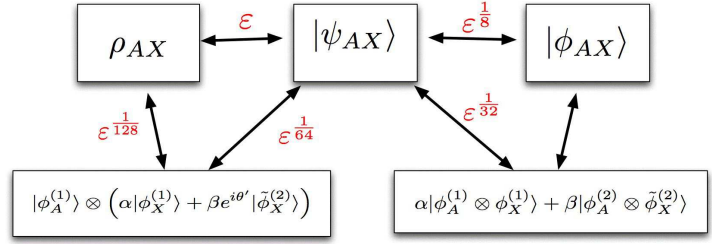}
\end{center}
\caption{Summary of the main inequalities
necessary to derive the threshold of Eq.~(\ref{varevare}).
The lines connecting the blocks represent
the fidelity
relation among the corresponding states (for each line we
reported the scaling of $1-F$ in terms of $\varepsilon$ by always considering
the worsth-case scenario). The starting point is the
inequality~(\ref{afa0000}) among $\rho_{AX}$ and the vectors
$|\psi_{AX}\rangle$. The final point is instead the
inequality~(\ref{totem10}) which connects $\rho_{AX}$ with
the vector $ \alpha  \; |\phi_A^{(1)} \otimes\phi_X^{(1)}\rangle
+ \beta \; e^{i \theta^{\prime}} \; | \phi_A^{(1)} \otimes \tilde{\phi}_X^{(2)} \rangle$.
 } \label{fig2}
\end{figure*}

For clarity we split the proof into three parts. In
part one
we show that the output states~(\ref{xmap})
associated with an orthonormal basis of the source system $S$ are
``close'' to a set of {factorizable} and  ``almost'' orthogonal
states of $ABC$. As in the perfect QSB case,  the finite size of $A$
will allows us to identify two elements of the selected basis of $S$
whose output images on $B$ and $C$ separately are described by
almost orthogonal vectors. In
part two we will use this
result to identify a two-dimensional subspace of $S$ whose image on
$A$ is ``almost'' constant. Finally, in part three
we will
show that the channel ${\cal N}$ yields ``good'' copies of such
subspace on $B$ and $C$. The threshold on $\varepsilon$ will follows by
direct comparison of the resulting fidelities with the optimal
cloning values~\cite{CLO2,OPTIMALCLO}. In Fig.~\ref{fig2} we
summarize the main passages of the derivation.

\subsubsection*{Proof of Theorem 2: Part one}\label{partone}
Assuming that $\varepsilon$-QSB from $S$ to $ABC$ is  possible
 for some given $\varepsilon\in (0,1]$,
consider an arbitrary orthonormal basis
$\{ |k_S\rangle\; ;  k = 1,\cdots , d_S \}$ of ${\cal H}_S$
and denote with
$|k_{AB} \rangle$ and $|k_{AC}\rangle$ their isometric representations
on $AB$ and $AC$, and with $|\phi_{A}^{(k)}\rangle$, $|\phi_{B}^{(k)}\rangle$ and
$|\phi_{C}^{(k)}\rangle$ the corresponding pure vectors of $A$, $B$
and $C$ which satisfy the conditions
of Lemma 1, e.g.
\begin{eqnarray} F(|k_{AX}\rangle; |\phi_A^{(k)}\otimes\phi_X^{(k)}\rangle) > 1
- \varepsilon^{\prime} \;, \label{jjj} \end{eqnarray} for $X=B,C$ and for all $k$.
Exploiting the orthogonality of $|k_S\rangle$ and using the triangle
inequality~(\ref{tri})  one can show that
  $\{|\phi^{(1)}_{A}\otimes\phi_B^{(1)} \rangle$, $\cdots$,
$|\phi^{(d_S)}_{A} \otimes \phi_B^{(d_S)}\rangle\}$ and
  $\{|\phi^{(1)}_{A}\otimes \phi_C^{(1)} \rangle$, $\cdots$,  $|\phi_{A}^{(d_S)} \otimes
\phi_C^{(d_S)} \rangle\}$
are two sets of ``almost'' orthogonal vectors, i.e.
\begin{eqnarray}
| {\langle} \phi^{(k^\prime)}_{A} \otimes \phi^{(k^\prime)}_X | \phi^{(k)}_{A}\otimes \phi_X^{(k)}\rangle | &=&
\label{hhh}
  | {\langle} \phi_A^{(k^\prime)} | \phi_A^{(k)}\rangle \;
{\langle} \phi_X^{(k^\prime)} | \phi_X^{(k)}\rangle | \nonumber \\
&&\qquad \qquad \qquad <  \varepsilon^{\prime\prime} \;,
\end{eqnarray}
for all $k\neq k^\prime$ and with $\varepsilon^{\prime\prime}$ being a small quantity depending
on $\varepsilon$.
In particular,
a rough estimation from Eq.~(\ref{goal1v})
gives $\varepsilon^{\prime\prime} = 2\sqrt{\varepsilon^{\prime}} + \varepsilon^{\prime}$, i.e.
$\varepsilon^{\prime\prime} \simeq 4.9 \;  \varepsilon^{1/4}$ for $X=B$
and
$\varepsilon^{\prime\prime} \simeq 7.1\;  \varepsilon^{1/16}$ for $X=C$.
Now we invoke the following result.
\newline

{\bf Lemma 2:} \emph{If $\{ |\phi^{(1)}\rangle, \cdots,
|\phi^{(m)}\rangle\}$ is a collection of $m > d$ unit vectors in a
Hilbert space ${\cal H}$ of dimension $d$ satisfying the condition}
\begin{eqnarray}
|\langle \phi^{(i)} | \phi^{(j)} \rangle | <  \xi \;,
\end{eqnarray}
\emph{for all $i\neq j$, then $\xi > \sqrt{\frac{m-d}{d(m-1)}}.$}
\newline

\emph{Proof:} Define the state $\rho \equiv \sum_{j=1}^m
|\phi^{(j)}\rangle\langle \phi^{(j)}|/m$ and use the equation
$\mbox{Tr}[ \rho^2 ] \geqslant 1/d$. $\blacksquare$
\newline

Lemma 2 requires  that for any collection of $m \geqslant d+1$ vectors of a
$d$-dimensional space there must be at least two whose scalar
product is greater than $1/d$. Since in our case $d_S \geqslant d_A
+1$ this implies that among the vectors
 $|\phi_A^{(1)}\rangle$, $\cdots$, $|\phi_A^{(d_S)} \rangle$ of ${\cal
H}_A$ there must exist at least  two elements (say $k=1$ and $k=2$)
which satisfy the inequality
\begin{eqnarray}
F(|\phi_A^{(1)}\rangle;|\phi_A^{(2)}\rangle) = |{\langle} \phi_A^{(2)} | \phi_A^{(1)} \rangle |^2
\geqslant 1/d_A^2\;.
 \label{ddd}
\end{eqnarray}
Taking then $\varepsilon^{\prime\prime} \leqslant {1}/{d_A}^2$
and replacing it into Eq.~(\ref{hhh}) we get
\begin{eqnarray}
 | \langle \phi_X^{(2)} | \phi_X^{(1)}\rangle | <  \sqrt{
\varepsilon^{\prime\prime}}  \;,  \label{titti}
\end{eqnarray}
which shows  that $|\phi_B^{(1)}\rangle$ and $| \phi_C^{(1)}\rangle$
 are almost orthogonal to
$| \phi_B^{(2)}\rangle$ and $| \phi_C^{(2)}\rangle$, respectively.

\subsubsection*{Proof of Theorem 2: Part two}\label{parttwo}

We now use Eq.~(\ref{titti})  to improve the inequality~(\ref{ddd})
showing that for small $\varepsilon$  the vectors $|\phi_A^{(1)}\rangle$ and
$|\phi_A^{(2)}\rangle$ are indeed close. To do so let us focus on
the two-dimensional subspace ${\cal H}_0$ of ${\cal H}_S$ formed by
the superpositions
\begin{eqnarray}
|\psi_S\rangle = \alpha |1_S\rangle + \beta |2_S\rangle \;, \label{vecotrs}
\end{eqnarray}
with $\alpha$ and $\beta$ complex amplitudes.
Their representations $|\psi_{AX}\rangle= V_{AXS} |\psi_S\rangle$ can
 be expressed as
\begin{eqnarray}
|\psi_{AX}\rangle  =
\alpha |1_{AX} \rangle + \beta |2_{AX}\rangle  \label{equi}\;,
\end{eqnarray}
where we used the linearity of $V_{AXS}$. Let us now define the
state
\begin{eqnarray}
|\tilde{\phi}_X^{(2)}\rangle \equiv e^{i \theta} \;
 \frac{|\phi_X^{(2)} \rangle -   \langle \phi_X^{(1)}
|\phi_X^{(2)} \rangle \;\; |\phi_X^{(1)} \rangle }{ \sqrt{ 1 -
| {\langle} \phi_X^{(1)} |
\phi_X^{(2)}\rangle|^2}}\;, \label{gigi}
\end{eqnarray}
with $\theta$ being a phase factor that will be defined later on.
The vector $|\tilde{\phi}_X^{(2)}\rangle$ is orthogonal to
$|\phi_X^{(1)}\rangle$ and, thanks to Eq.~(\ref{titti}), is close to
$|\phi_X^{(2)}\rangle$, i.e. $|\tilde{\phi}_X^{(2)}\rangle \approx
|\phi_X^{(2)}\rangle$. We will now show that, for all $|\psi_S\rangle
\in {\cal H}_0$, the representations~(\ref{equi}) can be faithfully
expressed in terms of superpositions of the orthonormal states
$|\phi_A^{(1)}\otimes \phi_X^{(1)}\rangle$ and $|\phi_A^{(2)}\otimes
\tilde{\phi}_X^{(2)}\rangle$. Indeed, taking into account the
inequalities Eq.~(\ref{goal1v}) and the conditions~(\ref{hhh}) and
(\ref{titti}), one can verify that there is a suitable choice of
phase $\theta$, such that
  for all $\alpha$ and $\beta$ we have
\begin{eqnarray}
F( |\psi_{AX}\rangle ; \alpha | \phi_A^{(1)} \otimes \phi_X^{(1)}\rangle
+ \beta | \phi_A^{(2)} \otimes \tilde{\phi}_X^{(2)} \rangle)
>
1 - \varepsilon^{\prime\prime\prime}  \label{totem}
\end{eqnarray}
with $\varepsilon^{\prime\prime\prime} = 3 \sqrt{\varepsilon^{\prime}} + \sqrt{\varepsilon^{\prime\prime}} + \varepsilon^{\prime\prime}$, i.e.
$\varepsilon^{\prime\prime\prime} \simeq 11.4 \; \varepsilon^{1/8}$ for $X=B$ and  $\varepsilon^{\prime\prime} \simeq 15.2 \;
\varepsilon^{1/32}$ for $X=C$.
According to Eq.~(\ref{goal1v}) the vector
$|\psi_{AX}\rangle$ should also be close to a separable state of the form
$|\phi_A \otimes \phi_X\rangle$.
By transitivity~(\ref{tripu}) and monotonicity~(\ref{fidodido})
 of the fidelity
we can hence derive the following inequalities
\begin{eqnarray} F(|\phi_A \otimes
&&\phi_X\rangle ; \alpha | \phi_A^{(1)} \otimes \phi_X^{(1)}\rangle +
\beta | \phi_A^{(2)} \otimes \tilde{\phi}_X^{(2)} \rangle) \nonumber \\
&&\qquad\qquad \qquad > 1 -\left(
\sqrt{\varepsilon^{\prime}} + \sqrt{\varepsilon^{\prime\prime\prime}} \right),
\label{yxy} \\
&&F(|\phi_A \rangle ;  |\alpha|^2 | \phi_A^{(1)} \rangle \langle \phi_A^{(1)} |
+ |\beta|^2  | \phi_A^{(2)} \rangle\langle \phi_A^{(2)} |) \nonumber \\
 &&\qquad\qquad \qquad
  > 1 -\left( \sqrt{\varepsilon^{\prime}} + \sqrt{\varepsilon^{\prime\prime\prime}} \right).
\label{yxy1} \end{eqnarray} In the limit of small $\varepsilon$, the latter
expression shows that, independently of the coefficient $\alpha$ and
$\beta$ the density matrix on the left hand side is close to a pure
state. This can only happen if the vectors $|\phi_A^{(1)}\rangle$ and
$|\phi_A^{(2)}\rangle$ are indeed almost parallel. The easiest way to
verify this is by computing the maximal eigenvalue
$\lambda_{\text{max}}$ of the density matrix $|\alpha|^2 |
\phi_A^{(1)} \rangle \langle \phi_A^{(1)} | + |\beta|^2  | \phi_A^{(2)}
\rangle\langle \phi_A^{(2)} |$ and by requiring that, for all choices of
$\alpha$ and $\beta$ it should be greater than or equal to the
fidelity associated with Eq.~(\ref{yxy1})
--- see  the convexity
property~(\ref{conv2}) of the fidelity.  This gives
\begin{eqnarray}
\lambda_{\text{max}} = \frac{1 + \sqrt{1 - 4 \; |\alpha \beta|^2 \;
\big[ 1- F(|\phi_A^{(1)}\rangle;|\phi_A^{(2)}\rangle)\big] \;,
} }{2}
\label{lambda}
\end{eqnarray}
and hence
\begin{eqnarray} \label{lambda1}
F(|\phi_A^{(1)}\rangle;|\phi_A^{(2)}\rangle) &>&
1 -4 ( \sqrt{\varepsilon^{\prime}} + \sqrt{\varepsilon^{\prime\prime\prime}}) \nonumber \\
&>& 1 - 19.4 \; \varepsilon^{1/16}
\;,
\end{eqnarray}
where the values of $\varepsilon^{\prime}$ and $\varepsilon^{\prime\prime\prime}$ of the case $X=B$ has
been employed to get the best scaling.

\subsubsection*{Proof of Theorem 3: Part three}\label{partthree}

%%%%%%%%%%%%%%%%%%%%%%%%%%%%%%%%%%%%%%%%%%%%%%%%%%%
%%%%%%%%%%%%%%%%%%%%%%%%%%%%%%%%%%%%%%%%%%%%%%%%

The idea is now to use Eq.~(\ref{lambda1}) together with
 the transitivity
and monotonicity conditions of the fidelity
to show that for all
 $|\psi_S\rangle$ of Eq.~(\ref{vecotrs})
the reduced density matrices $\rho_{AX}$
are close  to $ | \phi_A^{(1)} \rangle  \otimes
\Big(\; \alpha  \; | \phi_X^{(1)}\rangle \label{totem1}
+ \beta \; e^{i \theta^{\prime}} \; | \tilde{\phi}_X^{(2)} \rangle \; \Big)$
with the constant $\theta^{\prime}$ accounting for the relative phase
between $|\phi_A^{(1)}\rangle$ and $|\phi_A^{(2)}\rangle$.
Indeed we first notice that
\begin{eqnarray}
&&F\Big(\alpha |\phi_A^{(1)}\otimes \phi_X^{(1)}\rangle +
\beta |\phi_A^{(2)}\otimes \tilde{\phi}_X^{(2)}\rangle ; \nonumber \\
&& \qquad \qquad \qquad  \alpha  \; |\phi_A^{(1)} \otimes\phi_X^{(1)}\rangle \label{totem110}
+ \beta \; e^{i \theta^{\prime}} \; | \phi_A^{(1)} \otimes \tilde{\phi}_X^{(2)} \rangle \Big)\nonumber \\
&& \qquad \qquad \quad \geqslant  F(|\phi_A^{(1)}\rangle;|\phi_A^{(2)}\rangle)
> 1 - 19.4 \; \varepsilon^{1/16} \;.
\end{eqnarray} Exploiting the triangle inequality~(\ref{tripu}) twice we can
thus use Eqs.~(\ref{afa0000}) and (\ref{totem}) to show that \begin{eqnarray}
&&F\Big(\rho_{AX};
 \alpha  \; |\phi_A^{(1)} \otimes\phi_X^{(1)}\rangle \label{totem10}
+ \beta \; e^{i \theta^{\prime}} \; | \phi_A^{(1)} \otimes \tilde{\phi}_X^{(2)} \rangle \Big)
\nonumber \\
&&\qquad \qquad \qquad \qquad > 1 -   \varepsilon_{\text{iv}} \;,
\end{eqnarray} where for $X=B$ one has $\varepsilon_{\text{iv}}= 3.8 \;\varepsilon^{1/64}$
while for $X=C$ one has $\varepsilon_{\text{iv}}= 3.9 \;\varepsilon^{1/128}$.
The monotonicity property of the fidelity can then be invoked to
verify that the reduced density matrices $\rho_X$ are close to the
vector $\alpha  \; | \phi_X^{(1)}\rangle \label{totem3} + \beta \; e^{i
\theta^{\prime}} \; | \tilde{\phi}_X^{(2)} \rangle  = W_{XS} |\psi_S\rangle$, where
similarly to Eq.~(\ref{xxx32}), $W_{XS}$ is an isomorphism from
${\cal H}_0\in {\cal H}_S$ to ${\cal H}_X$ which maps $W_{XS}|1_S\rangle
= |\phi_X^{(1)}\rangle$, $W_{XS}|2_S\rangle = e^{i\theta^{\prime}}
|\tilde{\phi}_X^{(2)}\rangle$, i.e. \begin{eqnarray}\label{water1} F(\rho_X;
W_{XS}|\psi_S\rangle) > 1 -\varepsilon_{\text{iv}} \;. \end{eqnarray}

The above expression shows  that the QSB channel ${\cal N}$ produces
output states~(\ref{xmap}) whose reduced density matrices on $B$ and
$C$ are approximated  copies of the states $|\psi_S\rangle$ of the
two-dimensional subspace ${\cal H}_0\in {\cal H}_S$. The resulting
transformation \begin{eqnarray} |\psi_S\rangle {\longrightarrow} \left\{
\begin{array}{l}
\rho_{B} = \mbox{Tr}_C [ {\cal N}(|\psi_S\rangle\langle\psi_S|)] \\ \\
\rho_{C} = \mbox{Tr}_B [ {\cal N}(|\psi_S\rangle\langle\psi_S|)]\;,
\end{array}\right.
\end{eqnarray}
is indeed an (approximated) $1\rightarrow 2$
quantum cloner. Accordingly if $\varepsilon$-QSB could be realized
for arbitrarily small $\varepsilon$,  then
the fidelity of such copies with the input states
would  become arbitrarily close to one. This however is prevented by the
fact that the fidelities of {\em any}
$1\rightarrow 2$ cloning devices are bounded from above
by the value $5/6$~\cite{CLO2,OPTIMALCLO}.
By comparing this with  Eq.~(\ref{water1}),
and by taking into account the condition
$\varepsilon^{\prime\prime} \leqslant 1/d^2_A$ introduced in Eq.~(\ref{titti}),
we get the threshold  of Eq.~(\ref{varevare}):
for values of $\varepsilon$ smaller than such $\varepsilon_0$ we can hence conclude that
$\varepsilon$-QSB maps cannot be implemented.

This complete the proof of Theorem 2. $\blacksquare$

\section{Conclusions}\label{sec:conc}
Quantum Broadcasting protocols have been generalized to
include the possibility of producing output copies on partially overlapping systems.
In this context we have shown that
perfect Quantum Shared Broadcasting is  possible if and only if
 the overlap among the output systems is sufficiently large to include
all possible input states.
We have also analyzed the case of imperfect copies
proving the existence
of a finite upper bound $\varepsilon_0$
on the achievable fidelities below which no approximated QSB can be performed
when $d_S > d_A$.
Since the derivation of such threshold has been obtained by
imposing only some of (but not all) the necessary conditions
on the output states of the channels the reported value for $\varepsilon_0$
is probably not optimal. The characterization of the ultimate value for $\varepsilon_0$ and
its application in the context of quantum capacity characterization is currently under
investigation.
\newline
%\begin{acknowledgements}
The authors acknowledge support from the Quantum Information
research program of Centro di Ricerca Matematica Ennio De Giorgi of
Scuola Normale Superiore. A.H. acknowledges support of the RFBR
grant 06-01-00164-a and hospitality of V.G. during his visit to
Pisa.
%\end{acknowledgements}

\end{document}